\documentclass[10pt,a4paper,conference,compsocconf]{IEEEtran}
\usepackage{cite}
\usepackage[pdftex]{graphicx}
\usepackage[cmex10]{amsmath}
\usepackage{array}
\usepackage{mdwmath}
\usepackage{mdwtab}
\usepackage{eqparbox}

\usepackage{subcaption}
\usepackage{stfloats}
\usepackage{url}
\usepackage[table,dvipsnames]{xcolor}
\let\labelindent\relax
\usepackage{amssymb,pgfplots,enumitem,csquotes,booktabs,multirow,tabularx}
\usepackage[vlined,linesnumbered,ruled,resetcount]{algorithm2e}
\pgfplotsset{compat=newest} 
\usepackage{hyperref,cleveref}
\renewcommand{\b}{\bfseries}
\newcommand{\qt}{\enquote}
\newcommand{\N}{\mathbb N}
\renewcommand{\l}{\left(}
\renewcommand{\r}{\right)}

\usepackage[commentmarkup=uwave,final]{changes} 

\definecolor{scolor}{rgb}{0.733,0.843,0.890}
\definecolor{oocolor}{rgb}{0.783,0.893,0.940}
\newcommand{\bottomline}{\noalign{\global\arrayrulewidth=0.5mm}\arrayrulecolor{scolor}\hline}
\newcommand{\rowlight}{\rowcolor{oocolor}}
\newcolumntype{L}{>{\hspace*{-0.6\tabcolsep}}l}
\newcolumntype{R}{r<{\hspace*{-0.6\tabcolsep}}}
\newcolumntype{C}{c<{\hspace*{-0.6\tabcolsep}}}

\renewenvironment{thebibliography}[1]{
  \begin{oldthebibliography}{#1}
    \setlength{\itemsep}{0em}
    \setlength{\parskip}{0em}
}
{
  \end{oldthebibliography}
}

\begin{document}

\title{Transaction Fees on a Honeymoon: Ethereum's EIP-1559 One Month Later}



\author{\IEEEauthorblockN{Dani\"el Reijsbergen$^{1}$, Shyam Sridhar$^{2}$, Barnab\'e Monnot$^{2}$, Stefanos Leonardos$^{1}$,  Stratis Skoulakis$^{1}$,  Georgios Piliouras$^{1}$}
\IEEEauthorblockN{$^{1}$Singapore University of Technology and Design (SUTD),~$^{2}$Ethereum Foundation \\
daniel\_reijsbergen,stefanos\_leonardos,efstratios,georgios@sutd.edu.sg,~shyam.sridhar,barnabe.monnot@ethereum.org}
}

\maketitle

\begin{abstract}
Ethereum Improvement Proposal (EIP) 1559 was recently implemented to transform Ethereum's transaction fee market. EIP-1559 utilizes an algorithmic update rule with a constant learning rate to estimate a \emph{base fee}. The base fee reflects prevailing network conditions and hence provides a more reliable oracle for current gas prices.\par
Using on-chain data from the period after its launch, we evaluate the impact of EIP-1559 on the user experience and market performance. Our empirical findings suggest that although EIP-1559 achieves its goals \emph{on average}, short-term behavior is marked by intense, chaotic oscillations in block sizes (as predicted by our recent theoretical dynamical system analysis~\cite{leonardos2021dynamical}) and slow adjustments during periods of demand bursts (e.g., NFT drops). Both phenomena lead to unwanted inter-block variability in mining rewards. To address this issue, we propose an alternative base fee adjustment rule in which the learning rate varies according to an \emph{additive increase, multiplicative decrease (AIMD) update scheme}. Our simulations show that the latter  \emph{robustly  outperforms the  EIP-1559 protocol}  under various demand scenarios. These results provide evidence that \emph{variable learning rate} mechanisms may constitute a promising alternative to the default EIP-1559-based format and contribute to the ongoing discussion on the design of more efficient transaction fee markets.
\end{abstract}

\section{Introduction}\label{sec:intro}

Transaction fees have been an integral part of the emerging blockchain economies and the topic of heated debates in the blockchain community \cite{Eas19,ferreira2021dynamic,Rou21}. Ethereum, the  second largest blockchain platform in terms of market value (as of September 6th, 2021), recently launched its long-awaited \emph{London hard fork} which implements -- among other protocol upgrades -- Ethereum Improvement Proposal (EIP) 1559, which aims to radically transform its transaction fee market \cite{Con19,Rob21}.\par
In Ethereum's original fee market, users selected a \textit{gas price} that indicated how much they were willing to pay for their transactions to be included on the blockchain. This mechanism behaved like a (generalized) first-price auction in which miners, the \qt{auctioneers}, allocated the available transaction slots in each block to the highest paying users or \qt{bidders}. Unfortunately, it also shared all of a first-price auction's drawbacks \cite{Bute21,Mas01}: untruthful bidding, excessive intra- and inter-block variation of transaction fees, and (over-)bidding competition for faster inclusion. \par
The EIP-1559 transaction market reform was put forward to address these shortcomings. After a period of consultation and implementation, EIP-1559 came into effect on August 5, 12:33:42 PM UTC, at block 12,965,000 \cite{EthF21b,EthF21a}. Its main goals are to allow for more flexibility during changing market conditions by aiming for a long-term average target block size (half-full blocks) instead of blocks that are consistently full, and to make transaction fees more predictable (but not necessarily lower).\par
To achieve these goals, EIP-1559 sets a \textit{base fee} that is algorithmically updated after each block to reflect the prevailing market demand. Critical in these updates is a hard-coded learning rate (currently set at $12.5\%$) which calibrates the size of the adjustment according to the current block size. Unlike the traditional formats in which all fees were transferred from users to miners, the base fee is now \emph{burnt} \cite{karantias2020proof}, i.e., permanently removed from the total supply of Ethereum tokens (ether or ETH). Alongside newly minted tokens, miners are compensated for their work by an additional \textit{tip} set by the users, typically much lower than the base fee.\medskip

\paragraph{Motivation and Contributions}
Due to Ethere\-um's prominence in the blockchain ecosystem, EIP-1559 naturally attracted widespread academic and community attention even before its launch. In \cite{roughgarden2020transaction,Rou21}, its economic properties were studied, proving that it is \textit{incentive compatible} in the sense that it is rational for users to follow a simple bidding strategy whenever the current base fee is \added{comparable to market demand, i.e., does not under-price the limited supply of gas in a block}. Meanwhile, \cite{monnot2020ethereums,leonardos2021dynamical} analyzed its dynamics under different demand scenarios and \cite{vitalik2021amm,ferreira2021dynamic} initiated the
discussion about alternative (EIP-1559-like) transaction fee market mechanisms.\par
In this paper, we contribute to this growing literature on EIP-1559. We use blockchain data from the first month after EIP-1559's launch -- the \qt{\emph{honeymoon}} period during which legacy transactions are still accepted -- to evaluate EIP-1559's impact on user experience and network performance. We use this empirical data to simulate and argue about the performance of EIP-1559 under different learning rate configurations. Our contributions can be summarized as follows.
\begin{itemize}[leftmargin=*]
    \item We find that the current learning rate leads to slow adjustments during periods of demand peaks (e.g., NFT \cite{kugler2021non} drops) and to intense oscillations in block sizes during periods of stable demand.
    \item Both slow adjustments and oscillations have a negative effect on the user experience and, more importantly, on mining rewards. In particular, they can potentially generate strategic incentives for miners and hence undermine the stability of the mechanism.
    \item We then use the empirical data to run simulations with different learning rates. Our results confirm the intuition that higher (lower) learning rates perform well (poorly) during demand peaks and poorly (well) during periods of demand stability. On the positive side, all EIP-1559 variations achieve (in fact, slightly overshoot) the target block size (50\%) on average.
    \item Based on the above, we propose an alternative mechanism that utilizes an additive increase, multiplicative decrease (AIMD) update scheme to implement a variable learning rate. Our empirical results suggest that the AIMD mechanism combines the best of both worlds: it selects a low learning rate during periods of stable demand and reacts quickly during demand peaks to reflect increasing gas prices. The AIMD not only achieves the target block size on average (half-full blocks) but also considerably dampens the oscillations between almost full and almost empty blocks, which leads to more stable mining rewards and a more uniform user experience.
\end{itemize}
While our empirical results do not prove the optimality of the proposed AIMD mechanism under arbitrary conditions, they provide evidence that variable learning rates may considerably improve upon the current EIP-1559 format and may thus constitute a promising direction for future work in the ongoing discussion about EIP-1559-based transaction fee mechanisms \cite{vitalik2021amm,ferreira2021dynamic}. The code for our experiments is publicly available at \url{https://github.com/daniel-sutd/transaction-fees-on-a-honeymoon}.\medskip

\paragraph{Outline} The rest of the paper is structured as follows. In \Cref{sec:pre}, we present the main elements of Ethereum's blockchain protocol and EIP-1559. In \Cref{sec:data}, we describe our dataset and in \Cref{sec:simulations} our simulation environment.  \Cref{sec:eval} contains the evaluation of the simulation results. \Cref{sec:conclusions} concludes the paper with questions and directions for future work.

\section{Preliminaries}\label{sec:pre}

\subsection{Ethereum}\label{sub:ethereum}

Ethereum has a prominent position in the growing block\-chain ecosystem, as the first platform to support decentralized, self-executing pieces of software known as \textit{smart contracts} \cite{buterin2014next}.\medskip

\paragraph{Block Creation} The creation and execution of smart contracts on the Ethereum blockchain is achieved through \textit{transactions} which are ordered and grouped into \emph{blocks}. Whenever a new smart contract is deployed or a function of an existing contract is called, the operation is embedded in a transaction. The transaction is then broadcast to consensus nodes -- which in Ethereum's current \emph{proof-of-work} mechanism are called \textit{miners} -- through a peer-to-peer network \cite{wood2014ethereum}. The miners vie for the right to create new blocks for which they earn \emph{block creation rewards} \cite{But19}. Eventually, a new block that includes a set of new, previously pending transactions is created and broadcast to the nodes of the peer-to-peer network, who then update their view of the blockchain. Valid blocks that are not propagated across the network quickly enough can be  superseded by a competing block -- the losing blocks are called \emph{uncles}.\medskip

\paragraph{Transaction Fees}
The maximum size of the blocks that encapsulate these transactions is limited as per the specifications of the protocol and is measured in terms of the total amount of computational effort required for executing all transactions in that block. It is chosen as a trade-off between 1) higher throughput and 2) making it possible for more low-power nodes to participate. Thus, there is a demand for block space among users that wish to get their transactions included in the blockchain, which is supplied by the miners who create the blocks. To find a balance between this supply and demand, users must pay a \emph{transaction fee} to the miners for including their transaction in a block. This provides a second form of protocol-supported payment for the miners, alongside the block creation rewards.\medskip

\paragraph{Ethereum Gas}
The unit that measures the amount of computational effort required to execute transactions on the Ethereum blockchain is referred to as \emph{gas}. For example,  creation of a new smart contract may require $500,000$ units of gas, but a simple transfer of ETH tokens typically requires only $21,000$ units of gas. The use of gas units (or simply gas) allowed for a uniform implementation of transaction fees in Ethereum's original fee market design: for each transaction, the user would specify a \emph{gas price} that indicates the fee paid by the user per unit of gas used to execute the transaction, and a \emph{gas limit} that indicates the maximum amount of gas that the transaction is able to consume.\medskip

\paragraph{Ethereum's Legacy Fee Market} Ethereum's le\-gacy fee market mechanism, thus, closely resembled that of a \emph{(generalized) first-price auction}. Users would publish a transaction with a fee, miners would choose the highest-paying transactions, and each user would pay what they bid. The higher the user was willing to pay in transaction fees, the higher their chances of having their transactions included in a block. However, such auctions are well-known to be highly inefficient, leading to extreme fee volatility and frequent over-payments \cite{Mas01,Nis07}. This is especially observed when the network's transaction load is high, e.g., due to new project launches or periods of token price volatility. To remedy these shortcomings, EIP-1559 was proposed \cite{Bute21}. 

\subsection{EIP-1559 Transaction Fee Market}\label{sub:eip}

EIP-1559 is an upgrade on Ethereum's design which aims to address the above challenges by reforming the transaction fee market.
With its introduction on August 05, 2021 on Ethereum's mainnet, the maximum block size was doubled from roughly 15 million to $T=30$ million gas. However, the long-term average block size that the system targets is half that limit, namely $T/2$.\medskip

\paragraph{Base Fee} The main element of EIP-1559 to achieve this target is a dynamically adjusted \emph{base fee}, $b_t$, that is updated after every block, indexed by block height $t>0$, according to the equation
\begin{equation}\label{eq:basefee}
b_{t+1}=b_t \l 1+d\cdot\frac{G_t-T/2}{T/2}\r.
\end{equation}
Here, $d$ denotes the \emph{learning rate} (or \emph{adjustment parameter}), currently set at $d=0.125$, and $G_t$ the \added{total gas used by} transactions that were included in block $t>0$. Intuitively, if $G_t>T/2$, i.e., if more transactions than targeted are included in a block, then the base fee increases. The increase is scaled by $d$ and is proportional to the amount by which the current block load $G_t$ exceeds the target $T/2$. Similarly, if $G_t<T/2$, then the base fee decreases, whereas if $G_t=T/2$, then the base fee remains unchanged.\medskip
\paragraph{Bids, Tips, and Effective Gas Prices}
According to EIP-1559, instead of a single gas price, users now submit two parameters, $(f,p)$, as shown in \Cref{tab:eip}.

\begin{table}[!htb]
\centering
\setlength{\tabcolsep}{4pt}
\renewcommand{\arraystretch}{1.2}
\begin{tabularx}{\linewidth}{LX}
\bottomline
\rowlight
\b Parameter & \b Description \\
\bottomline
$f$: max fee & maximum amount \added{per  gas unit} that the user is willing to pay for their transaction to be included \\
$p$: max priority fee & maximum tip \added{per gas unit} that the user is willing to pay to the miner who includes their transaction
\\\\[-0.35cm]
\bottomline
\end{tabularx}
\caption{EIP-1559 user bid parameters.}
\label{tab:eip}
\end{table}
If the max fee that a user is willing to pay for their transaction is less than the algorithmically computed base fee, i.e., if $f<b_t$, then the transaction cannot be included in a block. Conversely, if $f \ge b_t$, then the miners can include the transaction and hence earn the so-called \textit{miner's tip}, which is calculated as
$\text{miner's tip}=\min{\{f-b_t,p\}}$
In practice, if the miner's tip is very small (less than 2 Gwei \added{per gas unit, where 1 Gwei = $10^{-9}$ ETH}), it may be beneficial for miners not to include the transaction. Larger blocks are harder to transmit over the network, making it easier for a competing block to reach other nodes faster, causing the miner to lose out on the block creation rewards and fees (\qt{uncle risk}).\par
Importantly (and in sharp contrast to the original fee market), the base fees are not transferred to miners: instead they are \emph{burnt} and are permanently removed from the total supply of ETH.\footnote{As of September 3rd, 2021, over 180,000 ETH were burned through EIP-1559 transactions (source: \href{https://watchtheburn.com/}{watchtheburn.com}).}
The sum of the base fee and tip fee taken per unit gas is said to be the \emph{effective gas price} and is equivalent to the gas price that a user would have paid to include their transaction before EIP-1559. \medskip

\paragraph{Legacy Transactions} EIP-1559 was introduc\-ed in a backward-compatible manner allowing both \emph{legacy} and native EIP-1559 transaction formats to be included. However, all transactions included after block 12,965,000 are subject to the base fee, regardless of their format. In particular, when a legacy transaction is submitted to the network, the max priority fee, $p$, and the max fee, $f$, are set by default equal to the gas price specified by the user. The miner then earns $f - b_t$ if the transaction is included in their block, possibly leading to overpayment at inclusion, compared to the same transaction following the simple EIP-1559 bidding strategy. With an EIP-1559 transaction, the user places a cap on the miner's tip by setting a max priority fee, whereas in a legacy transaction all the remaining fees, after subtracting the base fee, go to the miner.
\section{Data}\label{sec:data}

To analyze the impact of EIP-1559 during the honeymoon period, we use publicly available data recorded on the Ethereum blockchain. Our dataset covers a period starting from block $l=12935000$, which was mined on 3 August 2021, 18:43, until block $u=13079999$, which was mined on 23 August, 06:38. EIP-1559 came into effect in block 12965000, which was mined on 5 August 2021, 12:33 (the mentioned times are all in UTC). Our data set is described in detail in \Cref{tab:data} and visualized in the next paragraphs, which a focus on both \qt{long-term} trends (i.e., observations from the entire study period) and \qt{short-term} behavior (i.e., observations regarding sudden demand peaks or other transient phenomena).


\begin{table}[!htb]
\centering
\setlength{\tabcolsep}{4pt}
\renewcommand{\arraystretch}{1.3}
\begin{tabularx}{\linewidth}{LX}
\bottomline
\rowlight
\b Parameter & \b Description \\
\bottomline
$x\in \N$ & transaction, identified by its hash\\
$t_x \in \N$ & height of the block in which $x$ was included\\
$g_t \in [0,1]$ & \textit{relative} block size of the block at height $t$, i.e., the total used gas $G_t$ divided by the gas limit $T$\\
$b_t \in \N$ & base fee in the block at height $t$\\
$\tau_x \in \{0,1\}$ & type of $x$, which is \qt{$0$} for legacy transactions and \qt{$1$} for EIP-1559 transactions\\
$f_x \in \N$ & gas price of $x$ for legacy transactions, and the max fee of $x$ for EIP-1559 transactions\\
$p_x \in \N$ & priority fee of $x$ (for legacy transactions this is set equal to $f_x$)\\
$u_x \in \N$ & gas limit of transaction $x$ (used as a proxy for the gas that was actually used by the transaction)\\
$X(t) \subset \N$ & set of all transactions included in block at height $t$ \\\\[-0.35cm]
\bottomline
\end{tabularx}
\caption{Transaction information in the dataset.\vspace{-0.3cm}}
\label{tab:data}
\end{table}
\subsection{Long-term Trends}\label{sub:long}
\newcommand{\figsize}{0.685\linewidth}
\newcommand{\dualfigsize}{0.48\linewidth}

\begin{figure}[t!]
\centering
\subfloat[][Gas price]{\hspace*{-0.8cm}\includegraphics[width=\figsize]{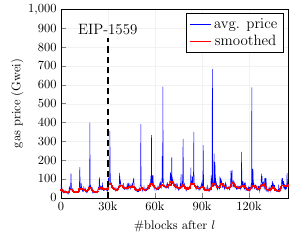}\label{fig:sim_price_stable}}\\[0.15cm]
\subfloat[][Fraction of EIP-1559 transactions]{\hspace*{-0.7cm}\includegraphics[width=\figsize]{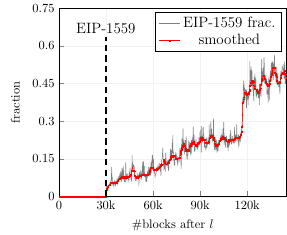}\label{fig:eip_1559_fraction}}
\caption{Top: average (blue) and smoothed averaged (red) gas prices over time. Bottom: average fraction (gray) and smoothed averaged fraction (red) of EIP-1559 transactions over time.}
\label{fig:average_gas_price}
\end{figure}

\paragraph{Average Gas Prices} First, we study the \emph{average block gas prices} during the sample period. To compute this metric over time, we first compute the average gas price $\hat{f}_t$ for each block at height $t$, where each transaction $x$ is weighted by its gas limit $u_x$:
\[\hat{f}_t = \l \sum\nolimits_{x \in X(t)} u_x f_x\r/\l\sum\nolimits_{x \in X(t)} u_x\r.\]
We then divide the set $\{l, l+1,\ldots,u\}$ into $1000$ batches of size $w=\frac{u-l+1}{1000}$, and compute the average $ \hat{f}^*_i$ for each batch $i \in \{1,\ldots,1000\}$ as
\begin{equation}\label{eq:batch}
        \hat{f}^*_i = \frac{1}{w} \sum\nolimits_{t=l+(i-1)w}^{l+iw}  \hat{f}_t.
\end{equation}
The result is displayed in \Cref{fig:average_gas_price}, which shows the evolution of the average block gas, $\hat{f}^*_i$.

We observe that the (average) gas prices exhibit sharp bursts. We have found that these bursts typically correspond to so-called \qt{NFT drops}, in which a creator (or project) announces the release of a limited number of Non-Fungible Tokens (NFTs) \cite{kugler2021non}. This leads to a demand surge where purchasers put in high bids to be included first by miners. Other, lower-scale fluctuations are due to diurnal variations in Ethereum's usage over different geographic locations. In \Cref{sub:short}, we return to the base fee and gas price dynamics and examine them in the locality of periods of both sharp demand increases and of relatively stable demand. \medskip

\paragraph{Smoothed Average Gas Prices} To dampen the effect of the sharp bursts on the long-term trend of gas prices, we apply a median smoothing filter to $\hat{f}^*_i$, which is unaffected by large outliers unlike the moving average. To compute the filtered value $m(\hat{f}^*_i,\eta)$ of $\hat{f}^*_i$, where $\eta$ is the half-width of the filter, we first determine the set
\[V(\hat{f}^*_i,\eta) = \{\hat{f}^*_j \in \N : j\geq l, j \leq u, |j-i| \leq \eta \}.\]
We then compute the median $m_i = m(\hat{f}^*_i,\eta)$ as the value such that at least $50\%$ of the elements in $V(\hat{f}^*_i,\eta)$ are greater or equal than $m_i$ and at least $50\%$ of the elements in $V(\hat{f}^*_i,\eta)$ are smaller or equal than $m_i$. The result with half-width $\eta=10$ is displayed by the red line in \Cref{fig:average_gas_price}.\medskip

\paragraph{Fraction of EIP-1559 Transactions}

We apply the same methodology as above to the weighted fraction, $\hat{\tau}_h$, of EIP-1559 transactions (versus legacy transactions) which for the block at height $t$ is given by
\[\hat{\tau}_t = \l\sum\nolimits_{x\in X(t)} u_x \tau_x\r/\l\sum\nolimits_{x \in X(t)} u_x\r.
\]
The result is displayed in \Cref{fig:eip_1559_fraction}. The gray line represents $\hat{\tau}^*_i$ averaged, as in equation \eqref{eq:batch}, over 1000 batches and the red line represents the corresponding smoothed values. The $i$-th value in the plot corresponds to the middle block of the $i$-th batch for $i\in\{1,2,\dots,1000\}$.

We observe that EIP-1559 adoption increases steadily within our sampling period to around $60\%$ at the end of August. The sharp increase around block 13051500 (8 August) corresponds to the (actual date of) adoption of EIP-1559 by the popular Ethereum wallet MetaMask. After that point, we observe strong diurnal fluctuations as a result of differences in wallet usage over different geographic locations.

\subsection{Short-term Behavior}\label{sub:short}

We now take a more detailed view at the short-term behavior of the base fee and gas price dynamics during periods of both sharp demand increases (e.g., NFT drops) and of relative demand stability.\medskip

\paragraph{Demand peaks}

\begin{figure*}[!tb]
\centering
\subfloat[][Burst period prices]{
\includegraphics[width=0.326\linewidth]{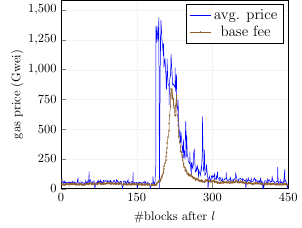}
\label{fig:burst}}
\subfloat[][Stable period prices]{
\includegraphics[width=0.313\linewidth]{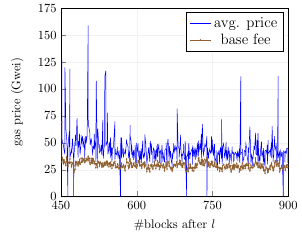}
\label{fig:stable}}
\subfloat[][Stable period sizes]{
\includegraphics[width=0.31\linewidth]{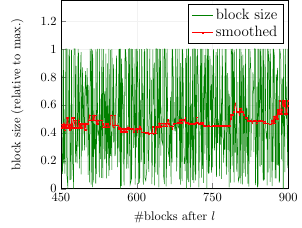}
\label{fig:stable_size}}
\caption{(a) Average gas prices (blue) and base fees (brown) per block in the volatile period after block 13025550. (b) Average gas prices (blue) and base fee per block (red) in Gwei, in the period of relatively stable demand that directly followed the burst of (a). (c) Relative block sizes, $\hat{g}_t$, in the same period of relatively stable demand as in (b).}
\end{figure*}

\Cref{fig:burst} zooms in on one of the price bursts that are visible in \Cref{fig:average_gas_price}, namely the one that occurs around block 13025775 that was mined on August 14, 2021, at 09:35. The displayed period covers 450 blocks between blocks 13025550 and 13026000.

Before the burst, average gas prices (depicted by the blue line without marks) fluctuate between 30 and 60 Gwei, with several short bursts in which the average fee jumps to around 120 Gwei before returning to around 45 Gwei, and several empty blocks (in which the average fee is set to $0$). The average gas price then shoots from 40.82 Gwei in block 13025736 to 207.33  and ultimately to 1360.02 Gwei within the next two blocks.\par
To compare, the miner of block 13025736 earns a total of 2.019 ETH: 2 ETH for the block creation reward, plus 0.137 in transaction fees minus 0.118 in burnt fees. In this case, as is a typical for blocks outside price bursts, the transaction fees are small compared to the block rewards \cite{carlsten2016instability}. However, the miners of blocks 13025737 and 13025738 earn 3.472 ETH and 45.646 ETH, respectively. In the latter case, the earnings from transaction fees are more than 20 times greater than the block creation reward! \par
\added{Concerning the base fee (depicted by the brown line with marks), we observe that shortly after the price burst it increases rapidly until it reaches a level slightly below the average gas price. After that point, it starts to decline together with gas prices which gradually return to normal. In sharp contrast to the above, block 13025768 is ultimately far from target, netting 0.844 ETH in transaction fees for 20\% block utilisation (10\% of the block gas limit). Comparing this revenue with that of the next miner 13025769, who received 4.868 ETH for 180\% block utilisation (90\% of the block gas limit), 13025768 misses out on some share of these fees. With a uniform process generating the transaction arrival time, this share may be estimated to 2 ETH (an equal split between 0.844 and 4.868 ETH), worth around $\$6,500$ USD at the time. These points demonstrate an instability in inter-block mining rewards under EIP-1559 that may require closer monitoring.}\medskip

\paragraph{Demand Stability}

\Cref{fig:stable} displays the gas price and base fee dynamics for the 450-block period of relatively stable demand that starts immediately after the burst of \Cref{fig:burst}. During this period, both the gas price and the base fee remain roughly stable: the former at around 40 Gwei and the latter slightly below 30 Gwei. However, despite the fact that the base fee remains within a relatively narrow range of 10 Gwei, i.e., between 25 and 35 Gwei, block sizes fluctuate wildly between the two extremes (empty to full and vice versa). This can be seen in \Cref{fig:stable_size}, where the green line (without marks) displays the relative block size as a number between $0$ (empty) and $1$ (full). This is exactly the behavior predicted in \cite{leonardos2021dynamical,monnot2020ethereums}, and can be attributed to the discrete nature of the updates and the size of $d$. \par
However, it is interesting to observe that block sizes do, in fact, seem to reach (or, actually, only slightly overshoot) the target block size \emph{on average}. As shown in \Cref{fig:stable_size}, the median filter with half-width 30 applied to $\hat{g}_t$ (red line with marks) is indeed centered around $0.5$ (half-full blocks). Providing theoretical reasons for or empirically testing this observation over longer periods of data is an interesting direction for future work.
\section{Simulations: Learning Rate in EIP-1559}\label{sec:simulations}

As established by the short-term analysis in the previous section, there are two opposing effects concerning the learning rate $d$. In particular, the default $d$ (set at 0.125) is \emph{too low} during periods of short bursts (during which the base fee lags in catching up to actual demand) but is also \emph{too high} during periods of relative stability (where it induces intense oscillations in block sizes despite keeping the base fee relatively stable). While long-term trends suggest that block-sizes achieve the target of half-full blocks \emph{on average}, these fluctuations have significant effects on miners' earnings and thus, potentially, on protocol stability.\par
As such, we investigate the impact of choosing a different $d$ (constant or variable) on the network. To do so, we run a series of simulations in which we observe the base fee after transaction bids are drawn randomly.

\subsection{Constant Learning Rate}\label{sub:constant}
The main challenge in simulating \added{(i.e., replaying using a stochastic model)} the evolution of the transaction fee market for different values of $d$ (or alternative mechanisms) is to draw accurate estimates for the actual demand and the users' valuations from the available data. The difficulty stems from the fact that observable bids 
-- i.e., those included on the blockchain -- 
do not necessarily reflect users' true valuations, but merely their valuations/expectations about the current and future base fees and their urgency to get their transaction included. Furthermore, they do not include low-value bids that were not included.\medskip

\paragraph{Behavioral Model}
As such, we make some assumptions -- justified by our current data and Ethereum's previous simulations, see \cite{Rob21,leonardos2021dynamical,monnot2020ethereums} -- and leave a deeper investigation of bids, e.g., with questionnaires to users, as an open challenge for future research. Accordingly, we model the number of users in each time slot using a Poisson demand with arrival rate $\lambda$ and a distribution of valuations, $F$. As mentioned above, both $\lambda$ and $F$ are subject to change over time -- e.g., due to the effects of weekends and weekdays, day/night cycles in different regions, sudden bursts due to popular contracts and NFT launches -- and therefore challenging to infer from the available data.\medskip 
\paragraph{Execution Model} Given a Poisson distribution of transaction arrivals (i.e., demand) with time-dependent arrival rate $\lambda_t$ and valuation distributions $F_t$, the simulator executes the steps of \Cref{alg:execution} in each time slot $t$.

\begin{algorithm}
\caption{Simulation of EIP-1559 Updates}
\label{alg:execution}
\KwIn{Demand parameters $\lambda_t$, $\hat{f}_t$, base fee $b_{t-1}$ at block height $t-1$ and learning rate $d$.}
\KwOut{base fee at block height $t$.}
$n_t \gets \textsf{draw\_num\_transactions}(\lambda_t)$\\
$m'_t \gets \textsf{draw\_transactions}(n_t, \hat{f}_t)$\\
$m_t \gets \textsf{m\_update}(m_{t-1},m'_t)$\\
$B_t \gets \textsf{B\_update}(m_t)$\\
$g_t \gets |B_t|/T$\\
$b_t \gets \textsf{b\_update}(b_{t-1},d,|B_t|)$

\end{algorithm}

\Cref{alg:execution} first calls $\textsf{draw\_num\_transactions}(\lambda_t)$, which pseudo-randomly draws a value $n_t$ from the Poisson distribution with mean $\lambda_t \equiv 3$ (adopted from \cite{leonardos2021dynamical}). The function $\textsf{draw\_transactions}(n_t, \hat{f}_t)$ then draws $n_t$ samples from $F_t$ that represent the valuations of the transactions created at time $t$, given the observed median-smoothed average gas price in the block at height $t$, denoted by $\hat{f}_t$. These values are then stored in the set $m'_{t}$. We have observed that a mixture of uniformly distributed and Pareto-distributed random samples, with means that depend on the smoothed average fees, results in a good correspondence with the observed values' distribution. \added{We then draw the valuations of $2.75n_t$ transactions from a uniform distribution near $0.75 \hat{f}_t$, and draw the remaining $0.25n_t$ valuations from a Pareto distribution with shape $1.35$ and scale $\frac{1}{10} \hat{f}_t$. The reasoning behind this modelling choice is to divide users into two broad classes. The first class consists of \emph{regular} users for whom the delay before the transaction appears on the blockchain is relatively unimportant, and who simply bid a value that (they believe) ensures that their transaction is eventually included. The second class consists of \emph{high-urgency} users who make high bids to ensure that their transaction is included rapidly, and whose valuations more closely resemble a heavy-tailed distribution. The parameters were chosen to give a good correspondence with the studied dataset, but we have found that minor variations did not change the overall conclusions of our evaluation, cf. \Cref{tab:simulations}.}

After having drawn $m'_t$, \Cref{alg:execution} combines the transactions in $m'_t$ with the process $m_t$, which represents the \emph{mempool}, i.e., the set of transactions that are pending to be included in the blockchain. 
The function $\textsf{m\_update}(m,m')$ returns a new mempool given an existing mempool $m$ and a set $m'$ of new transactions. 
In our simulations, transactions remain in the mempool until they are included.
The function $\textsf{B\_update}(m)$ is then executed by the miner. $\textsf{B\_update}(m)$ returns a vector $B$ with the max fees of the transactions that have been included in the new block given the mempool $m$. We denote the length of $B$ as $|B|$. Finally, the function $\textsf{b\_update}(b,d,|B|)$ returns the new base fee as a function of the current base fee $b$, the learning rate $d$ and the block size $|B|$.

\subsection{Variable Learning Rate: AIMD Updates}\label{sub:aimd}

To address the issue that the current learning rate appears too high in stable-demand periods and too low in periods with demand peaks, we also propose and simulate a mechanism with a variable learning rate. In particular, motivated by the effectiveness of such schemes to control internet congestion \cite{yang2000general}, we study an \emph{additive increase and multiplicative decrease} (AIMD) update scheme for the learning rate $d$.

\begin{table}[!htb]
\centering
\setlength{\tabcolsep}{6pt}
\renewcommand{\arraystretch}{1.2}
\begin{tabularx}{\linewidth}{LX}
\bottomline
\rowlight
\b Parameter & \b Description \\
\bottomline
$g_{\textsc{AVG}}$ & average relative block-size of previous blocks\\
$n \in \N$ & nr.\ of previous blocks over which $g_{\textsc{AVG}}$ is computed.\\
$\alpha > 0$ & size of an additive increase \\
$\beta \in [0,1]$ & factor of a multiplicative decrease\\
$\gamma \in [0,\frac{1}{2}]$ & threshold for an additive increase\\
$d_{\min}, d_{\max}$ & minimum and maximum values for $d$.
\\\\[-0.35cm]
\bottomline
\end{tabularx}
\caption{Parameters of the AIMD algorithm.}
\label{tab:aimd}
\end{table}

According to the AIMD scheme, the learning rate is adjusted with respect to a parameter, $g_{\textsc{AVG}}$, that keeps track of the average relative block sizes ($g_t=G_t/T$, cf. \Cref{tab:data}) during some pre-specified rolling window. The intuition is the following: when $g_{\textsc{AVG}}$ is close to the target block-size, $0.5$, then the base fee is close to the right value, so the algorithm reduces the learning rate to reduce the size of the oscillations. By contrast, if $g_{\textsc{AVG}}$ is too small or high, then the base fee is apparently far away from its equilibrium value, and the algorithm increases the learning rate.\medskip 

\paragraph{Execution Model (AIMD)}
The AIMD update scheme uses the parameters in \Cref{tab:aimd}. It then replaces line 6 in \Cref{alg:execution} with the following two lines
\setlength{\interspacetitleruled}{2pt}%
\begin{algorithm}
\caption{Modified Alg. \ref{alg:execution} with AIMD updates}
\label{alg:line6}
\setcounter{AlgoLine}{5}
$d_t \gets \textsf{d\_update}(d_{t-1},g_{t-n+1},\ldots,g_{t})$\\
$b_t \gets \textsf{b\_update}(b_{t-1}, d_t, |B_t|)$
\end{algorithm}    
which update the learning rate parameter $d$ according to \Cref{alg:aimd} (following the intuition described above):

\setlength{\interspacetitleruled}{2pt}%
\begin{algorithm}
\caption{AIMD update routine $\textsf{d\_update}$.}\label{alg:aimd}
\KwIn{Learning rate, $d_{t}$, window $n$, block-sizes $g_{t-n+1},\ldots,g_{t}$, update parameters $\alpha,\beta,\gamma$.}
\KwOut{Learning rate, $d_{t+1}$.}

$g_{\textsc{AVG}} \gets \frac1n  \sum^t_{t-n+1} g_i$
\vspace{0.1cm}\\
\uIf{$g_{\textsc{AVG}} < \gamma \textnormal{\textbf{ or }} g_{\textsc{AVG}}> 1 - \gamma$}{
$d_{\textsc{new}} \gets \min\{d_{\max}, \alpha + d\}$
}
\uElse{
$d_{\textsc{new}} \gets \max\{d_{\min}, \beta d\}$
}
\Return $d_{\textsc{new}}$
\end{algorithm}
\section{Evaluation}\label{sec:eval}
We now proceed to compare the different update me\-chanisms described in the previous section. We present simulations of the standard EIP-1559 update rule with constant step sizes at 
$d=0.125$ (default), $d=0.25$ and $d=0.0625$ which are representative of high and low rates, respectively, and of the AIMD (variable step size) update rule with parameters $n=8$, $\alpha = 0.025,\beta=0.95,\gamma=0.25, d_{\min} = 0.0125$ and $d_{\max}=1$. Note that $n$ is selected to be low enough to reflect recent market conditions, but high enough to reduce its sensitivity to temporal fluctuations in demand.

\subsection{Performance Metrics}\label{sub:metrics}
To compare the performance of the different mechanisms, we use the two metrics described in \Cref{tab:metrics}.
\begin{table}[!htb]
\centering
\setlength{\tabcolsep}{8pt}
\renewcommand{\arraystretch}{1.2}
\begin{tabularx}{\linewidth}{LX}
    \bottomline
\rowlight
\b Parameter & \b Description \\\\[-0.3cm]
\bottomline
$\hat{g}$ & long-term average relative block size\\
$p_{g>0.95}$ & long-term fraction of blocks whose relative size is greater than $0.95$.
\\\\[-0.35cm]
\bottomline
\end{tabularx}
\caption{Performance metrics.}
\label{tab:metrics}
\end{table}

The first metric, $\hat{g}$, should be as close to $0.5$ as possible, as this is the target set by the miners.
\begin{figure*}[htb!]
     \centering
     \subfloat[][$d=0.0625$]{\includegraphics[width=0.24\textwidth]{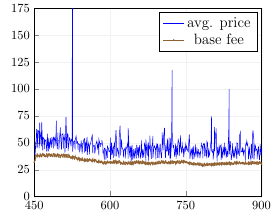}\label{fig:sim_price_stable_slow}}
     \subfloat[][$d=0.125$]{\includegraphics[width=0.24\textwidth]{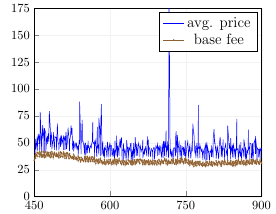}\label{fig:sim_price_stable_2}}
     \subfloat[][$d=0.25$]{\includegraphics[width=0.24\textwidth]{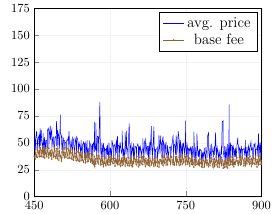}\label{fig:sim_price_stable_fast}}
     \subfloat[][AIMD]{\includegraphics[width=0.24\textwidth]{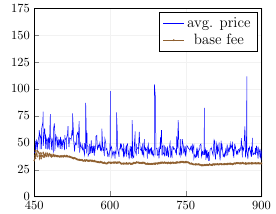}\label{fig:sim_price_stable_aimd}} 
     
     \subfloat[][$d=0.0625$]{\includegraphics[width=0.24\textwidth]{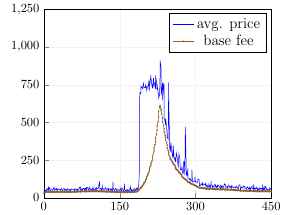}\label{fig:sim_price_burst_slow}}
     \subfloat[][$d=0.125$]{\includegraphics[width=0.24\textwidth]{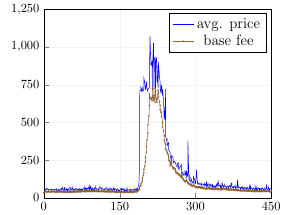}\label{fig:sim_price_burst}}
     \subfloat[][$d=0.25$]{\includegraphics[width=0.24\textwidth]{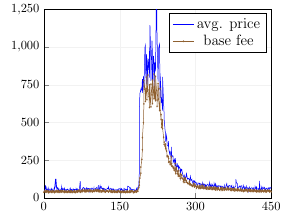}\label{fig:sim_price_burst_fast}}
     \subfloat[][AIMD]{\includegraphics[width=0.24\textwidth]{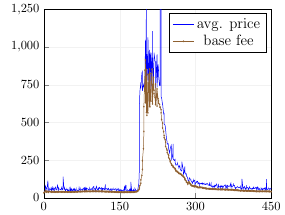}\label{fig:sim_price_burst_aimd}}
     
     \caption{Simulation of average gas prices in Gwei (blue lines) and base fees (red lines) in stable (top) and peak demand periods (bottom) of the EIP-1559 market under different learning rates. Unlike standard EIP-1559 mechanisms with slow or fast learning rates ($d=0.125$ being the default), the AIMD mechanism exhibits good performance in both scenarios.}
     \label{fig:simulations}
\end{figure*}
When $\hat{g}$ is (considerably) larger than $0.5$, propagation and execution time of new blocks increase, putting stress on the network of miners to validate all new items. Conversely, when $\hat{g}$ is (considerably) lower than $0.5$, network capacity is under-utilized. \added{
The situation where $\hat{g}$ is slightly below 0.5 more closely resembles the original target setting of the Ethereum network before  EIP-1559 with max block size of $T/2$ making such states  marginally preferable over ones where $\hat{g}$ is slightly above $0.5$.
} \par
The second metric, $p_{g>0.95}$ should be as low as possible, even if $\hat{g}$ approximates $0.5$ (e.g., due to full blocks counterbalancing empty blocks). \added{When block-sizes are consistently high, the transaction fee market approaches the behavior of a first-price auction, losing some of the advantages of EIP-1559 \cite{roughgarden2020transaction}}.\footnote{The reason that we do not consider $p_{g=1}$ is purely technical: since not all transaction have the same size, in some cases a block will only use, say, $99\%$ of its gas and yet it will still be considered \qt{full} in the sense that no other available transactions can be added to it.}

\subsection{Simulation Results}\label{sub:results}

The effect of choosing different learning rates in the standard EIP-1559 mechanism is displayed in \Cref{fig:simulations}. We visualize two scenarios: the stable period from \Cref{fig:stable} (top row), and the NFT drop from \Cref{fig:burst} (bottom row). The blue lines show the simulated demand which approximates the actual demand, cf. \Cref{fig:stable,fig:burst}.\par 
As intuitively expected, the oscillations of the base fee (brown line with marks) in the stable scenario decrease as the learning rate decreases. Similarly, in the NFT drop scenario, the fast learning rate mechanism, $d=0.25$, swiftly adapts to the changing situation, whereas the other two learning rate mechanisms generate long periods of discrepancies between the base fee and market demand. \par
\begin{figure*}
     \centering
     \captionsetup[subfigure]{aboveskip=0.5pt,belowskip=0.5pt}
     \begin{subfigure}[b]{0.495\textwidth}
     \centering
     \includegraphics[width=0.48\textwidth]{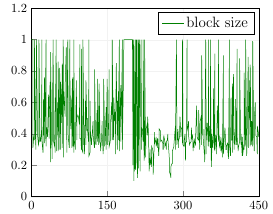}\label{fig:sim_g_burst_aimd}
     \includegraphics[width=0.48\textwidth]{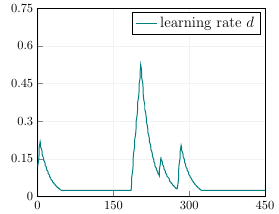}\label{fig:sim_d_burst_aimd}
     \caption{AIMD: demand peak}
     \end{subfigure}
     \begin{subfigure}[b]{0.495\textwidth}
        \centering
     \includegraphics[width=0.48\textwidth]{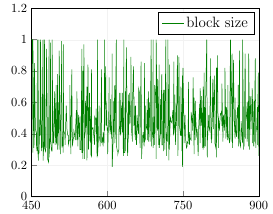}\label{fig:sim_g_stable_aimd}
     \includegraphics[width=0.48\textwidth]{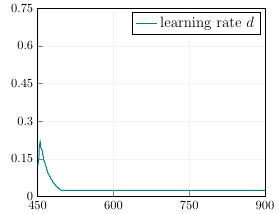}\label{fig:sim_d_stable_aimd}
     \caption{AIMD: stable demand}
     \end{subfigure}
     \caption{Block sizes and learning rate evolution under the variable learning rate AIMD scheme during the demand peak (first and second panel) and stable demand periods (third and fourth panels), respectively. AIMD adjusts the learning rate during the demand peak and uses a lower than default learning rate to dampen the oscillations (in particular, empty or almost empty blocks) during the stable demand period (cf. EIP-1559 performance in \Cref{fig:burst,fig:stable}.}\vspace{0.2cm}
     \label{fig:simulations_g_d}
\end{figure*}

The important takeaway, however, concerns the AIMD mechanism (panels in the last column), which turns out to share the advantages of both approaches. In particular, when the system is relatively stable, it selects a learning rate of $d=0.0125$ (lower than the current default of $d=0.125$) that increases swiftly during demand peaks. This results in low oscillations of the base fee during the stable period (top row) and a timely increase during the demand peak period (bottom row). To emphasize this, we have also displayed in \Cref{fig:simulations_g_d} the block sizes and the evolution of the learning rate $d$ for the AIMD scheme. We set $d=0.125$ as the initial value, and observe that in both scenarios the learning rate decreases to its minimum values, $d_{\min}=0.0125$. However, in the NFT drop scenario, the learning rate rapidly increases to over 0.45, and then decreases again when the base fee starts to reflect market conditions. In the stable scenario, we observe from \Cref{fig:sim_g_stable_aimd} that although there are still oscillations, they are noticeably less intense than in \Cref{fig:stable_size}. \par
To further quantify this performance, \Cref{tab:simulations} contains the results for both metrics, $\hat{g}$ and $p_{g > 0.95}$, over three scenarios: the stable period (450 blocks), the demand peak period (450 blocks), and our full dataset of blocks after EIP-1559 (115,000 blocks).
\newcommand{\gaussconfintv}[2]{\begin{tabular}{c} #1 \\[-0.125cm] \scalebox{0.7}{$\pm$#2} \end{tabular}}

\begin{table*}[!tb]
\centering
\setlength{\tabcolsep}{6pt}
\renewcommand{\arraystretch}{1.07}
\begin{tabular}{@{}clcc@{\hskip 0.4in}clcc@{\hskip 0.4in}clcc@{}} 
    \bottomline
    \rowlight
    &\hspace{-17pt}\b Stable periods & $\hat{g}$ & $p_{g > 0.95}$&
    &\hspace{-17pt}\b Demand peaks & $\hat{g}$ & $p_{g > 0.95}$&
    &\hspace{-17pt}\b Full dataset & $\hat{g}$ & $p_{g > 0.95}$
        \\\\[-0.3cm]
    \bottomline
    & empirical & 0.523 & 0.261 & 
    & empirical & 0.528 & 0.281 &
    & empirical & 0.523 & 0.244 \\[0.12cm]

    \multirow{6}{*}{\rotatebox{90}{simulation}}     & $d=0.0625$ &  \gaussconfintv{0.505}{0.000} & \gaussconfintv{0.150}{0.005} &
        \multirow{6}{*}{\rotatebox{90}{simulation}}  & $d=0.0625$ & \gaussconfintv{0.511}{0.000} &  \gaussconfintv{0.222}{0.004} &
    \multirow{6}{*}{\rotatebox{90}{simulation}}     & $d=0.0625$ & \gaussconfintv{0.505}{0.000} & \gaussconfintv{0.186}{0.000} \\

    & $d=0.125$  &  \gaussconfintv{0.510}{0.000} & \gaussconfintv{0.214}{0.004} &
    & $d=0.125$ & \gaussconfintv{0.513}{0.000} &  \gaussconfintv{0.232}{0.004} &
    &  $d=0.125$ & \gaussconfintv{0.510}{0.000} & \gaussconfintv{0.232}{0.000} \\

     & $d=0.25$ &  \gaussconfintv{0.520}{0.000} & \gaussconfintv{0.255}{0.002} &
     & $d=0.25$ & \gaussconfintv{0.522}{0.000} &  \gaussconfintv{0.265}{0.003} & 
     & $d=0.25$ & \gaussconfintv{0.520}{0.000} & \gaussconfintv{0.264}{0.000} \\

     & AIMD & {\bf \gaussconfintv{0.499}{0.001}} & {\bf \gaussconfintv{0.058}{0.003}} &
     & AIMD & {\bf \gaussconfintv{0.481}{0.002}} &  {\bf \gaussconfintv{0.126}{0.003}} & 
     & AIMD & {\bf \gaussconfintv{0.501}{0.000}} & {\bf \gaussconfintv{0.047}{0.000}} \\\\[-0.3cm]
     \bottomline
\end{tabular}
\caption{Empirical evaluation of the different EIP-1559 mechanisms with low, default ($d=0.125$), and high learning rates in terms of the performance metrics $\hat{g}$ (average block size) and $p_{g > 0.95}$ (probability that a block is more than $95\%$ full). The AIMD mechanism has the best performance in terms of $p_{g > 0.95}$ in both the stable and unstable scenarios, and across the entire post-EIP-1559 dataset (bold values).}\vspace{-0.2cm}
\label{tab:simulations}
\end{table*}
\added{In \Cref{tab:simulations}, below each point estimate of $\hat{g}$ and $p_{g > 0.95}$ we also display the half-width of a 95\% confidence interval for the corresponding value based on the normal distribution. For each entry in the table, we conducted $20$ simulation experiments -- despite the small sample size, we note that the variance is fairly low since each sampled estimate of $\hat{g}$ and $p_{g > 0.95}$ is itself an average over at least 450 blocks. For each scenario, we also display the \textit{empirical} result, i.e., the values of $\hat{g}$ and $p_{g > 0.95}$ that were observed in the data set.} \par
Importantly, the AIMD has a low number of full blocks in all three scenarios and hence exhibits the best performance in terms of $p_{g>0.95}$ by a wide margin when compared to the other methods. AIMD updates also achieve the desired average relative block size ($\hat{g}\approx 0.5$). Consistent with our empirical observations, all standard EIP-1559 updates also approximate, yet, consistently overshoot by a small margin, the target block-size on average, but perform poorly on the $p_{g>0.95}$ metric. This suggests the presence of more than 20\% almost full blocks counterbalanced by a similar amount of almost empty blocks.
\subsection{\added{Discussion}}
\added{Given the desirable properties showcased by the AIMD over the standard EIP-1559 update rules in the above simulations, it is natural to ask whether the AIMD update rule could actually fit in the current scope of EIP-1559. In particular, the default, more moderate EIP-1559 update that corresponds to a fixed $d=0.125$ aims to achieve the following goals: (i) generate more predictable fees against comparable market conditions, (ii) adapt fees to prevailing demand, (iii) fight inflation by burning excessive fees instead of transferring them to miners, and (iv) tame strategic behavior by protocol participants, both miners and users. In the following, we discuss how an AIMD update rule would perform along these axes.}
\paragraph{Predictability and Adaptation}
\added{Concerning the first two goals, there is a certain amount of trade-off between short-term fee predictability and swiftly responding to changing market conditions. However, the high-level goals of next-block inclusion and invariant block sizes tip the scale towards more responsive fee mechanisms.}
\paragraph{Excessive Fees} \added{Concerning the goal of controlling inflation, it is tempting to think that a slow-adapting mechanism may burn more fees by allowing full blocks during demand peaks. However, users' valuations are typically skewed towards higher values \cite{Eas19,Hub21} and higher fees at periods of increased demand can actually reap higher amounts of burned coins.}
\paragraph{Strategic Behavior}
\added{The fourth goal, that of mitigating strategic behavior by miners or users, is more intricate and merits a separate discussion. Apart from general attacks on the PoS consensus layer \cite{Neu21,Sch21}, there is an open debate on the vulnerabilities of the EIP-1559 fee market (in its original format) against exacerbated versions of existing or novel attack vectors \cite{But19}. One such example is the \emph{zero-base fee attack}, in which a coalition of miners who control 51\% of the network's resources collectively drive the base fee down to zero by continuously mining empty blocks. At that point the only determining factor for transaction inclusion are miner's tips, essentially moving the system back to the previous first price auction model. While the practical feasibility of such an attack is still unclear, it is a necessary to look at the effects of the more responsive AIMD update rule on miner's incentives via a more comprehensive threat model. We leave this as a direction for future work.}

\section{Conclusions \& Open Questions}\label{sec:conclusions}
Using on-chain data from the first month after its launch, we studied the early stages of Ethereum's EIP-1559 fee market upgrade. We found evidence that the current fixed learning rate -- the parameter that is used to adjust the base fee after each block based on network congestion -- achieves the target block-size on average (or rather, slightly overshoots it). However, it may have a considerable destabilizing effect on mining earnings -- via slow adjustments during demand bursts and intense oscillations in block occupancies during periods of stable demand -- and on the user experience. Such phenomena may lead to strategic behavior and ultimately pose a threat to the stability of the fee market.\par
Our simulations suggest that a promising direction to improve upon the current EIP-1559 format is the use of variable learning rates. In particular, we studied an additive increase, multiplicative decrease (AIMD) update scheme for the learning rate that resulted in superior performance over standard EIP-1559 formats across periods of both demand bursts and demand stability. \par
Given this range of results, a number of interesting open questions emerge. Specifically, it will be instructive to monitor the impact of EIP-1559 over longer time spans (i.e., beyond the \qt{honeymoon} period), to study different learning rate update schemes (e.g., multiplicative increase, additive decrease, or exponential updates) and to derive accurate estimates of Ethereum's recurring demand conditions to systematically optimize the parameters of the proposed AIMD scheme. \added{Finally, as mentioned above, it will be interesting to develop a strategic threat model to analyze how miners' and users' incentives change between the default EIP-1559 mechanism and the proposed AIMD update rule.}
Any progress in the above directions will contribute to the ongoing discussion about EIP-1559 and to the design of more efficient transaction fee mechanisms in general. 

\section*{Acknowledgment}
This research/project is supported in part by the National Research Foundation, Singapore under its AI Singapore Program (AISG Award No: AISG2-RP-2020-016), NRF 2018 Fellowship NRF-NRFF2018-07, NRF2019-NRF-ANR095 ALIAS grant, grant PIE-SGP-AI-2018-01, AME Programmatic Fund (Grant No. A20H6b0151) from the Agency for Science, Technology and Research (A*STAR) and the Ethereum Foundation. It is also supported by the National Research Foundation (NRF), Prime Minister's Office, Singapore, under its National Cybersecurity R\&D Programme and administered by the National Satellite of Excellence in Design Science and Technology for Secure Critical Infrastructure, Award No. NSoE DeST-SCI2019-0009.

\bibliographystyle{IEEEtran}
\bibliography{ref}

\end{document}